%% file: main.tex
\documentclass[camera]{jpaper}
\usepackage{cite}

\usepackage{siunitx}
\usepackage{amsmath}
\usepackage{listings}
\usepackage{graphicx}
\usepackage{subfig}
\usepackage{algorithm}
\usepackage[noend]{algpseudocode}	
\usepackage{stackengine}
\usepackage{tabularx}

\usepackage{algpseudocode}
\usepackage{titlesec}
\usepackage{siunitx}
\usepackage[all,defaultlines=3]{nowidow}
\algrenewcommand\algorithmiccomment[2][\normalsize]{{#1\hfill\(\triangleright\) #2}}

\titlespacing*{\section}{0pt}{3pt}{-1pt}
\titlespacing*{\subsection}{0pt}{3pt}{1pt}
\titlespacing*{\subsubsection}{0pt}{1pt}{0pt}

\usepackage{array}
\makeatletter
\let\MYcaption\@makecaption
\makeatother

\makeatletter
\let\@makecaption\MYcaption
\makeatother

\usepackage{dblfloatfix}
\usepackage[nolessnomore, italic]{mathastext}
\usepackage[T1]{fontenc}
\usepackage[usenames,dvipsnames,svgnames,table]{xcolor}
\usepackage{multirow}
\usepackage{hhline}
\usepackage[normalem]{ulem}
\usepackage{setspace}
\usepackage{indentfirst}
\usepackage{footmisc}
\usepackage{tabularx}
\usepackage[flushleft,para]{threeparttable}
\usepackage{tabularx,ragged2e}
\usepackage{cprotect}
\usepackage[keeplastbox]{flushend}

\usepackage[us,12hr]{datetime}
\usepackage{booktabs} 
\usepackage{mathtools}
\usepackage{xspace}

\newif\ifcameraready
\camerareadytrue

\ifcameraready
  \newcommand{\todo}[1]{}

\else
  \newcommand{\todo}[1][]{\textbf{\fcolorbox{black}{red}{\color{white}{TODO}}} \underline{$\overline{\hbox{\textcolor{red}{\emph{#1}}}}$}}

\fi

\definecolor{amber}{rgb}{1.0, 0.49, 0.0}
\definecolor{darkgreen}{rgb}{0.0, 0.2, 0.13}
\definecolor{darkbyzantium}{rgb}{0.36, 0.22, 0.33}
\definecolor{darkseagreen}{rgb}{0.56, 0.74, 0.56}
\definecolor{darkspringgreen}{rgb}{0.09, 0.45, 0.27}
\definecolor{dollarbill}{rgb}{0.52, 0.73, 0.4}

\newcommand{\affilETH}[0]{\textsuperscript{1}}
\newcommand{\affilVMWare}[0]{\textsuperscript{2}}
\newcommand{\affilPenState}[0]{\textsuperscript{3}}
\newcommand{\affilVU}[0]{\textsuperscript{4}}
\newcommand{\affilCMU}[0]{\textsuperscript{5}}

\newcommand{\clwb}{\texttt{clwb}\xspace}
\newcommand{\rntw}{\texttt{rntw}\xspace}
\newcommand{\rwtw}{\texttt{rwtw}\xspace}
\newcommand{\sfence}{\texttt{sfence}\xspace}
\newcommand{\rofence}{\texttt{rofence}\xspace}
\newcommand{\rdfence}{\texttt{rdfence}\xspace}

\newcommand{\rcommit}{\texttt{rcommit}\xspace}
\newcommand{\nosm}{NO-SM\xspace}
\newcommand{\smrc}{SM-RC\xspace}
\newcommand{\smob}{SM-OB\xspace}
\newcommand{\smdd}{SM-DD\xspace}
\newcommand{\etal}{et al.\xspace}

\begin{document}
%

\title{\setstretch{0.8}Enabling Efficient RDMA-based Synchronous Mirroring\\ of Persistent Memory Transactions}


\author{%
{Arash Tavakkol\affilETH}\quad%
{Aasheesh Kolli\affilVMWare$^,$\affilPenState}\quad%
{Stanko Novakovic\affilVMWare}\quad%
{Kaveh Razavi\affilETH$^,$\affilVU}\quad
{Juan\ G\'omez-Luna\affilETH}\\%
{Hasan Hassan\affilETH}\quad%
{Claude Barthels\affilETH}\quad%
{Yaohua Wang\affilETH}\quad%
{Mohammad Sadrosadati\affilETH}\quad%
{Saugata Ghose\affilCMU}\\%
{Ankit Singla\affilETH}\quad%
{Pratap Subrahmanyam\affilVMWare}\quad%
{Onur Mutlu\affilETH$^,$\affilCMU}\vspace{2pt}\\%
{\small\it\affilETH ETH Z{\"u}rich  \qquad \affilVMWare VMware \qquad \affilPenState The Pennsylvania State University \qquad \affilVU Vrije  Universiteit Amsterdam \qquad \affilCMU Carnegie Mellon University}%
\vspace{-5pt}%
}

\maketitle

\ifcameraready
  \pagenumbering{gobble}
\fi

\newcommand{\versionnum}[0]{7.1}

\fancyhead{}
\ifcameraready
 \thispagestyle{plain}
 \pagestyle{plain}
\else
 \fancyhead[C]{\textcolor{MidnightBlue}{\emph{Version \versionnum~---~\today, \ampmtime}}}
 \fancypagestyle{firststyle}
 {
   \fancyhead[C]{\textcolor{MidnightBlue}{\emph{Version \versionnum~---~\today, \ampmtime}}}
   \fancyfoot[C]{\thepage}
 }
 \thispagestyle{firststyle}
 \pagestyle{firststyle}
\fi

\setstretch{0.837}
\renewcommand{\footnotelayout}{\setstretch{0.9}}


\input{00.abstract}

\section{Introduction}
\input{00.intro}

\section{Background}
\input{01.background.tex}
\section{PM Transactions}
\input{02.transactions.tex}

\section{Memory-level SM}
\input{03.sm_rdma.tex}

\section{Replication Strategies}
\input{04.strategies.tex}

\section{Methodology}
\input{05.methodology.tex}

\section{Evaluation}
\input{06.real_applications.tex}

\section{Related Work}
\input{07.related_work.tex}

\section{Conclusion}
\input{08.conclusion.tex}

\SetTracking
 [ no ligatures = {f},
 outer kerning = {*,*} ]
 { encoding = * }
 { -40 } 

{

  \let\OLDthebibliography\thebibliography
  \renewcommand\thebibliography[1]{
    \OLDthebibliography{#1}
    \setlength{\parskip}{0pt}
    \setlength{\itemsep}{0pt}
  }
  \bibliographystyle{IEEEtranS}
  \bibliography{09.RDMA-SynchronousMirroring}
}

\end{document}

%% file: 00.abstract.tex
\begin{abstract}
Synchronous Mirroring (SM) is a standard approach to building highly-available and fault-tolerant enterprise storage systems.
%
SM ensures strong data consistency by maintaining multiple exact data replicas and synchronously propagating every update to all of them.
Such strong consistency provides fault tolerance guarantees and a simple programming model coveted by enterprise system designers.
For current storage devices, SM comes at modest performance overheads.  This is because performing both local and remote updates simultaneously is only marginally slower than performing just local updates, due to the relatively slow performance of accesses to storage (e.g., hard drives,
flash-based solid-state drives) in today's systems.
However, emerging persistent memory (or storage class memory) and ultra-low-latency network technologies necessitate a careful re-evaluation of the existing SM techniques, as these technologies present fundamentally different latency characteristics compared to their traditional counterparts. In addition to that, existing low-latency network technologies, such as \text{Remote Direct Memory Access (RDMA)}, provide limited ordering guarantees and do not provide durability guarantees necessary for SM.

To evaluate the performance implications of RDMA-based SM, we develop a rigorous testing framework that is based on emulated persistent memory.
Our testing framework makes use of two different tools: (i) a configurable microbenchmark and (ii) a modified version of the WHISPER benchmark suite, which comprises a set of common cloud applications, with support for SM over RDMA.
Using this framework, we find that recently proposed RDMA primitives,
such as remote commit, provide correctness guarantees, but do not take full advantage 
of the asynchronous nature of RDMA hardware. To this end, we propose new primitives enabling efficient and correct SM over RDMA, and use these primitives to develop two new techniques delivering high-performance SM 
of persistent memories. Overall, we find that our two SM designs outperform the remote commit based design by 1.8x and 2.9x, respectively.
\end{abstract}

%% file: 00.intro.tex
\label{section:introduction}

Enterprise applications require high-availability and durability guarantees. User data loss or service downtime can have significant economic impact or even lead to bankruptcy. 
For example, JournalSpace has been shut down due to a complete loss of the company's main database~\cite{journalspace}.
In August 2013, Amazon services stopped operating for a few minutes and the company lost millions of dollars in revenue~\cite{amazondowntime}. In such situations, it is of essence for any company to ensure that they can keep serving their users by relying on a different copy of the user data. To provide such uninterruptible business continuity, most companies maintain multiple copies of data in different failure domains.


Synchronous Mirroring (SM) is one of the most widely used approaches to building highly-available storage systems for the enterprise. Technologies like RAID 1, EMC MirrorView/S~\cite{emc}, VMware vSAN~\cite{vsan}, IBM Metro Mirror~\cite{ibm}, and NetApp SnapMirror~\cite{netapp} provide synchronous mirroring over short distances to ensure high availability and disaster
recovery. SM guarantees strong consistency of data by employing (at least) a pair of replica servers that host an exact copy of the application data. SM ensures that in the rare event of a failure, the application can rely on another active replica to continue seamless operation using the \emph{most recent version} of the data.
To provide such a guarantee, the underlying system has to propagate \emph{each} data update \emph{synchronously} to all of the replicas.
Unlike more weakly-consistent systems, such as asynchronous mirroring (AM),
SM (1)~guarantees zero data loss; and (2)~takes advantage of a simplified programming model, since
the programmer does not need to worry about reading an out-of-date replica.


Existing enterprise systems prefer to provide SM instead of weakly-consistent techniques such as AM, even though the weakly-consistent techniques lower performance overheads~\cite{emc, vsan, ibm, netapp}.  This is because in existing systems, 
the additional overhead of performing SM in place of techniques such as AM is relatively small~\cite{wadafekete11}.
The small overhead is a result of the relatively small latency difference between a local update
and a remote update today.  In systems that use off-the-shelf magnetic hard disk drives (HDDs) and
Ethernet controllers, waiting for a remote update via Ethernet does not take significantly longer
than a local update to the HDD.
While the relative cost of a remote update over Ethernet increases when the local system uses a solid-state
drive consisting of NAND flash memory, prior work~\cite{klimoviclitz17} has proposed domain-specific operating systems
and network stacks that greatly reduce the round-trip latency for a remote update.
As a result, the additional overhead of SM over AM remains insignificant in systems with SSDs.

Unfortunately, persistent memory technologies such as PCM, MRAM, ReRAM, and Intel's 3D XPoint~\cite{PMM-ISCA-2009,stt,reram,3dxpoint} make it much more
difficult to close the gap.  These persistent memories are expected to replace traditional storage devices (HDDs/NAND flash based SSDs)
in the near future~\cite{SNIA-White-Paper-2016,PMM-SOSP-2009,pelleychen14}, and are orders of magnitude faster than current storage media (100,000X faster than HDDs and 1000X faster than SSDs~\cite{SNIA-White-Paper-2016}).  Since local updates will take much less time with storage class memory, remote SM updates will become even more expensive relative to local updates. Without significant innovation on the remote update latency, it will become increasingly
more difficult to amortize and justify the additional overhead of SM over AM.


A promising approach to maintaining the viability of SM for persistent memory is
to make use of high-performance networks such as Infiniband~\cite{infiniband}, which are already becoming commonplace in modern datacenters~\cite{zhueran15, novakovicdaglis16}.
Prior works~\cite{RDMA-PMEM-SDC-2015,RDMA-PM-Requirements-HP-SDC-2014,Mojim-2015,SNIA-White-Paper-2016} have already proposed to combine fast Remote Direct Memory Access (RDMA) or Infiniband with the emerging
memories to improve the efficiency of SM, but these works do not perform a detailed study of how these remote access protocols affect the latency trade-offs between the remote access, local access, and application performance.

Our goal is to understand the performance implications of using RDMA to provide SM for storage class memory, which we refer to as \emph{memory-level SM} (SM hence forth), and propose new hardware primitives that will enable more efficient memory-level SM.
RDMA requires a specialized RDMA-enabled Network Interface Card (RNIC). Applications access the local RNIC directly from user-space to schedule read and write commands to the remote RNICs.
The end-to-end latency of RDMA operations depends on the complex interplay between the workload characteristics, processor architecture, and RNIC architecture. Finally, even
though RNIC sits on the PCIe bus, it is capable of writing and reading from the CPU cache
directly using Intel's Data Direct Input Output (DDIO) technology~\cite{ddio}. However, because the CPU cache resides outside of the persistence domain~\cite{intel_wq}, enforcing durability in RDMA-based SM is a challenge.

We develop a rigorous testing framework to evaluate memory-level SM using RDMA over an Infiniband network.  Our workloads consist of (1)~a configurable transaction microbenchmark and (2)~applications from the \emph{WHISPER}~\cite{nalliharia17} benchmark suite that have been extended with support for SM over RDMA.  Using our framework and workloads,
we find that an SM design that uses the recently proposed remote commit (\emph{rcommit}) primitive~\cite{rcommit} is unable to take full advantage of the asynchronous nature of the RDMA hardware.
We find that \emph{rcommit} is an expensive primitive because it is overloaded to provide both ordering and durability of SM updates.
And that significant performance gains can be made by developing separate primitives for ordering and durability and using only the appropriate primitive.
Similar insight has been used to improve the performance of single-node persistent memory systems~\cite{nalliharia17,pelleychen14} and file systems~\cite{chidambarampillai13}.

Inspired by our analysis, we design a new set of RDMA hardware primitives (i.e., \emph{verbs}): remote ordering fence, remote durability fence, remote write-through, and remote non-temporal writes. These new RDMA primitives aim to unleash the full potential of RDMA hardware and enable efficient memory-level SM. To showcase the new primitives, we propose two SM techniques, one based on ordered buffering, which leverages ordering, durability fences and remote
write-through; and another one based on non-temporal writes and the implicit ordering guarantees provided by modern RDMA hardware.

We use our framework to evaluate the effectiveness of our new SM techniques.
We find that the techniques provide much greater performance for memory-level SM than
remote commit based SM.  On average across our benchmarks, we find that our techniques
improve the program execution time by 1.8x and 2.9x and the throughput by 1.9x and 3.4x, respectively.
Thus, we argue that new primitives allowing for 
asynchronous updates, such as non-temporal remote writes, will be 
necessary for efficient SM over RDMA. 

We make the following contributions in this work:
\begin{itemize}
    
    \item We conduct a detailed study of RDMA-based synchronous mirroring of persistent memory using the existing remote commit verb. We find that
    remote commit verb does not
    take full advantage of the RDMA capabilities. 
    
    \item We identify that remote commit is inefficient as it is overloaded to provide both ordering and durability of SM updates. And that by developing separate primitives for ordering and durability we can significantly reduce the overheads of SM.

    \item We propose a new set of hardware primitives that are key to performing SM efficiently: remote ordering fence, remote durability fence, remote write-through, and remote non-temporal writes.
    
    \item We use our new primitives to develop two new SM techniques
    that our techniques significantly outperform RDMA-based SM using the remote commit verb.
    
    \item We design and implement a benchmark suite comprising of an RDMA microbenchmark and a version of the WHISPER benchmark suite that has been extended for memory-level RDMA-based SM. 
       
 
\end{itemize}

%% file: 01.background.tex
\label{section:background}


\subsection{Persistent Memories}
The emerging non-volatile memory technologies (e.g., PCM~\cite{PMM-ISCA-2009} and Intel 3D XPoint~\cite{NVM-Revolution-Intel-SDC-2015}) provide DRAM-like performance combined with disk-like durability~\cite{PMM-SOSP-2009,PMM-ISCA-2009,pelleychen14}.
These new memory technologies can be used to store and manipulate persistent data directly in-memory.
Such systems, referred to as {\em persistent memory systems}, have the potential to revolutionize how we manage persistent data and have sparked a wide variety of exciting research recently~\cite{izraelevitzmendes16,nawabizraelevitz17,venkataramantolia11,PMM-SOSP-2009,dulloor2015systems,PMM-SC-2011,volosnalli14,xuswanson16,arulrajperron16,pelleywenisch13,chatzistergioucintra15,aguilera2018remote,tavakkol2018mqsim,wangjohnson14,PMM-ICCD-2014,kollipelley16}.

Persistent software running on persistent memory systems requires in-memory data to be recoverable across system failures.
Ensuring recoverability requires programmers to have control over the order in which writes/updates reach the data in memory~\cite{pelleychen14,chidambarampillai13}.
However, in the presence of system failures (power outages, application/kernel crashes, etc.) today's systems do not provide any guarantees on the order in which writes reach persistent memory due to performance optimizations, like write back caching and memory controller reordering, which coalesce and reorder writes to persistent memory.
For example, consider adding a new node to a persistent linked list in memory.
The linked list can become inconsistent during an insertion operation if a power failure happens after the pointer to the new node has been persisted in memory, but the write that fills the new node with valid data did not persist in memory.
The recovery process of the linked list would observe a node with garbage values.
To ensure ordered updates, all the way to persistent memory, programmers have rely on a \emph{memory persistency model}~\cite{pelleychen14} and annotate their programs with appropriate low-level primitives, like Intel's $clwb$~\cite{intel_isa} instruction.

\subsection{Synchronous Mirroring}

To guarantee high availability and durability of business-critical and mission-critical data, system designers employ some form of data replication.
SM is a commonly used replication strategy where multiple replicas are kept in sync by propagating every update to all the replicas simultaneously, and is a feature on many commercial storage platforms~\cite{emc,vsan,ibm,netapp}.
In this paper, we limit ourselves to studying a variant of SM that involves only two replicas. 
All the updates originate at the primary server and are propagated to both local media and to the replica.
However, the insights gained from our work can be applied to a much broader class of replication strategies.

Storage-level SM has minimal performance overhead on the system, because storage write occupies significant portion of the latency of the entire persistent update operation (even the remote update)~\cite{maccormickmurphy04}.

For memory-level SM, to perform a persistent update, along with issuing a write to local persistent memory (i), the DMA engine fetches the data from the cache (ii) and sends a write request to the replica over the network (iii).
Upon receiving the request, the replica server performs a persistent memory write (iv) and then sends back an acknowledgement to the main server (v), completing the persistent update.
Since local persistent memory writes are much faster than remote writes (100s of ns vs 1000s of us~\cite{SNIA-White-Paper-2016}), the overheads of memory-level SM (over a baseline of just local updates) are prohibitive.
Storage-level SM (i.e., mirroring hard disks or SSDS) does not face the same problem because storage write occupies significant portion of the latency of the entire persistent update operation (even the remote update)~\cite{maccormickmurphy04}.
To make memory-level SM feasible, system designers need to employ ultra-low latency networks.
Next, we present an overview of RDMA, the ultra-low latency network protocol of choice in modern datacenters.


\subsection{RDMA}
\textit{Remote Direct Memory Access} (RDMA) is a mechanism enabling a user-level
program to directly access the memory of another computer, without interrupting the processors.
There are several existing networks that provide support for RDMA, of which the two most widely used are Infiniband~\cite{RDMA-Programming-Manual-2015} and an implementation using Converged Ethernet.
InfiniBand (IB)~\cite{RDMA-Programming-Manual-2015} is a complete network stack consisting of specialized transport, network, link, and physical layers. IB is commonly used in high-performance computing environments and is now making inroads into datacenters. IB offers low-latency and high-throughput communication, as it is
implemented entirely in hardware and allows for kernel bypass.

RDMA over InfiniBand requires special network gear, including a specialized switch, the links, and \emph{Host Channel Adapters (HCA)} or \emph{RDMA Network Interface Card (RNIC)}. For many RDMA applications it is important that the network can guarantee reliable delivery, meaning packets should not be dropped at any point in time. 
The RNIC implements the entire network
stack in hardware, allowing user-level programs to directly interact with
the controller and request remote operations. In existing RNIC implementation, the memory that is exposed to another server must be first \emph{registered} with the RNIC~\cite{RDMA-Programming-Manual-2015}. A portion of main memory that has been registered is referred to as a \emph{memory region}. During registration, the RNIC device driver pins the memory region in order to guarantee that this memory never gets swapped out from physical memory and thus is always accessible for 
RDMA operations. Registration also makes the virtual-to-physical address translation information (i.e., the page tables) accessible to the RNIC. A special copy of the page table entries that is designated for the RNIC is called \emph{Memory Translation Table} (MTT). MTT entries reside in memory and are buffered in the HCA cache. 


User applications initiate RDMA transfers by sending their requests directly to the RNIC using the \emph{verbs} API. There are two categories of verbs: \emph{messaging verbs}  and \emph{memory verbs}. These categories differ in terms of semantics and required inter-process synchronization.
In this work, we limit our study to memory verbs due to their lower overheads.
Memory verbs are called \emph{one-sided} operations since the execution of the remote read and write operations does not involve the CPU on the target machine. The lack of CPU overhead at the responder makes one-sided verbs attractive; they provide
the lowest latency and highest throughput among all verbs~\cite{HERD-SIGCOMM-2014}. In this work, we use one-sided RDMA writes to implement synchronous mirroring.

RNICs have been designed to be compatible with commodity servers while, at the same time, providing a low-overhead hardware-software interface. RNIC connects with the rest of the server components using the PCIe bus, where
the CPU communicates with the RNIC using PCIe commands. The RNIC uses the
PCIe bus to DMA to and from local memory. 
Most recently, Intel introduced the \emph{Data Direct I/O Access} (DDIO)~\cite{ddio} feature in its CPUs, allowing RNIC to read and write directly to the CPU cache, reducing 
the overhead of DMA controller invalidations, which used to be the key mechanism enabling coherent memory accesses in RDMA.

%% file: 02.transactions.tex
\label{sec:transactions}

\begin{figure}[t] \centering
\includegraphics[width=0.98\linewidth]{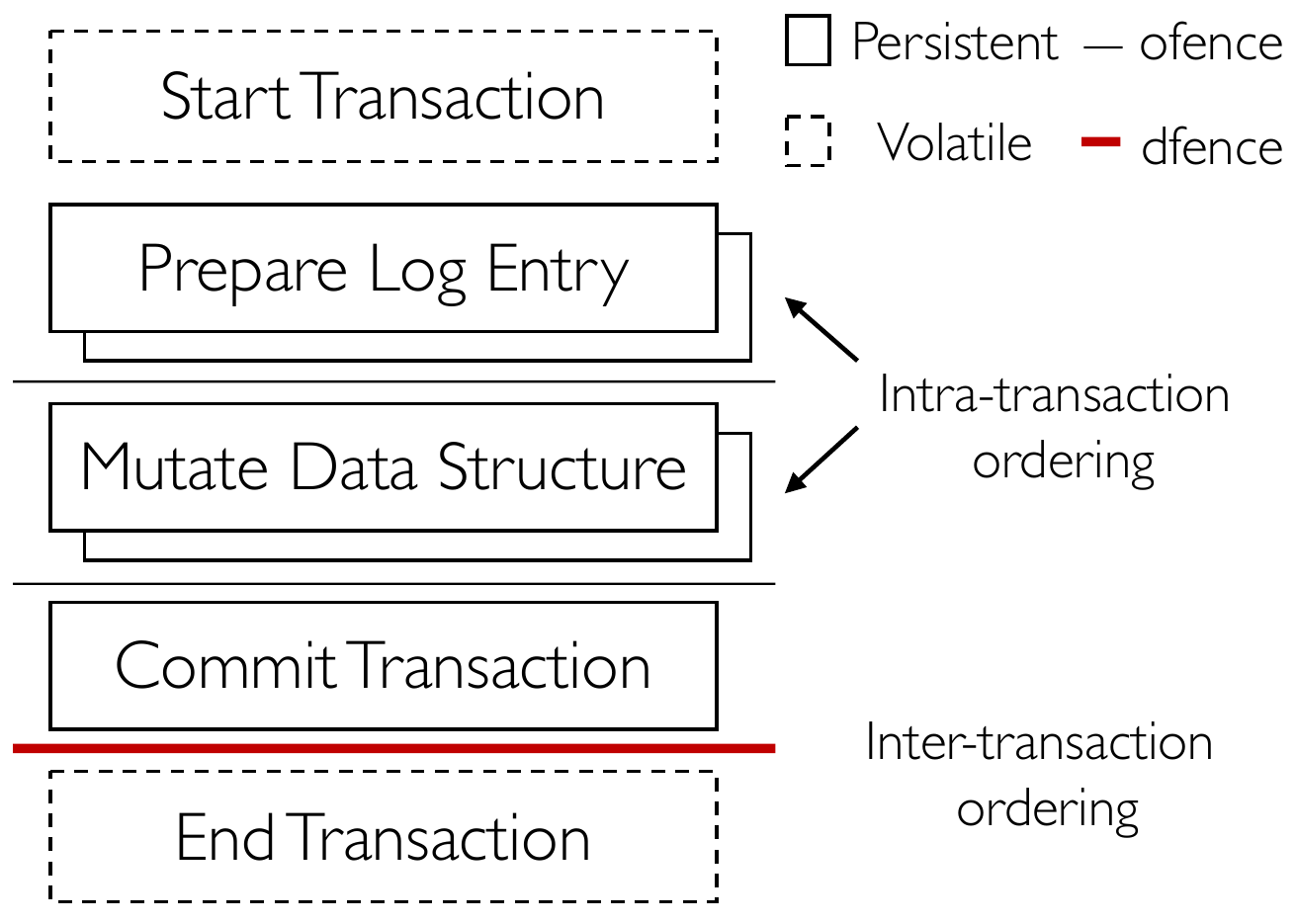}%
\caption{A typical undo-logging transaction~\cite{kollipelley16,nalliharia17} using ordering and durability fences to enforce intra-transaction ordering guarantees and inter-transaction (or end of transaction) durability guarantees. }
\label{fig:transaction}
\end{figure}

In this work, we assume that the applications modifying persistent memory use a storage transactions~\cite{kollipelley16, nalliharia17} for failure-atomicity.
Figure~\ref{fig:transaction} shows a typical transaction that uses undo logging to achieve failure-atomicity.
Storage transactions do not provide concurrency control and programmers have to use a separate mechanism (e.g., locks) to achieve concurrency control.
Next, we describe the guarantees provided by storage transactions and the primitives necessary to provide these guarantees.

\textbf{Guarantee-1: Failure-atomicity} This guarantee provided by the transaction system developer states that either all the updates from a transaction or none of the updates from a transaction are visible during recovery after a system failure.
The typical way to achieve failure-atomicity is to employ some form of logging.
For example, with undo logging (Fig~\ref{fig:transaction}), first a log entry (back up copy) of the data being modified is prepared, then the actual data is modified, and finally the log entry 
(back up copy) is invalidated to commit the transaction.
If during the execution of the transaction a failure occurs, the log entry is used to roll back the partial updates and provide the illusion of failure-atomicity.

To achieve failure-atomicity, a primitive that provides ``ordering'' guarantees between different steps of a transaction is mandatory.
Without such a primitive, failure-atomicity \emph{cannot} be provided.
For example, in the undo logging transaction, the programmer must have a way to ensure that the updates from the PrepareLogEntry step persist before any from MutateDataStructure persist.
As long as such an ordering primitive is provided to the programmer, the programmer can develop software crash consistency mechanisms (not just undo/redo logging) to provide failure-atomicity.
Hence forth, we refer to portions of a transaction between two successive ordering points as an \textit{epoch}.
So, a transaction may contain multiple epochs, all updates within on epoch may persist concurrently, however, updates from successive epochs must persist in order.

\textbf{Guarantee-2: Durability} While failure-atomicity provides a guarantee over what parts of a transaction may persist, the durability guarantee describes when a transaction must persist.
The end of a transaction is considered a ``durability'' point, i.e., after the completion of the transaction, all the updates of the transaction must be persisted.
The importance of a durability point is best understood in its contrast with an ordering point.
A ordering point simply enforces the order in which two sets of updates (before and after the ordering point) must persist.
It does not specify any constraints on when the sets of updates an be considered persisted, that is the responsibility of a durability point.
A durability point, once executed past, guarantees that all prior updates have persisted.

\textbf{Primitive-1: Ordering Fence} The first primitive, necessary to achieve failure-atomicity, is one that provides ordering guarantees.
We refer to a primitive that provides ordering guarantees as an ordering fence (\texttt{ofence}).

\textbf{Primitive-2: Durability Fence} The second primitive is one that provides durability guarantees.
We refer to a primitive that provides durability guarantees as a durability fence (\texttt{dfence}).

It is important to note that while some architectural proposals~\cite{pelleychen14,nalliharia17} provide separate primitives to provide ordering an durability guarantees, some architectures overload the ordering and durability functionality on the same primitive (for example, \texttt{sfence} in x86.~\cite{kollirosen16})

%% file: 03.sm_rdma.tex
\label{section:smrdma}
\subsection{SM over RDMA}
\label{section:smguarantees}

\begin{figure*}[t] \centering
\includegraphics[width=0.85\linewidth]{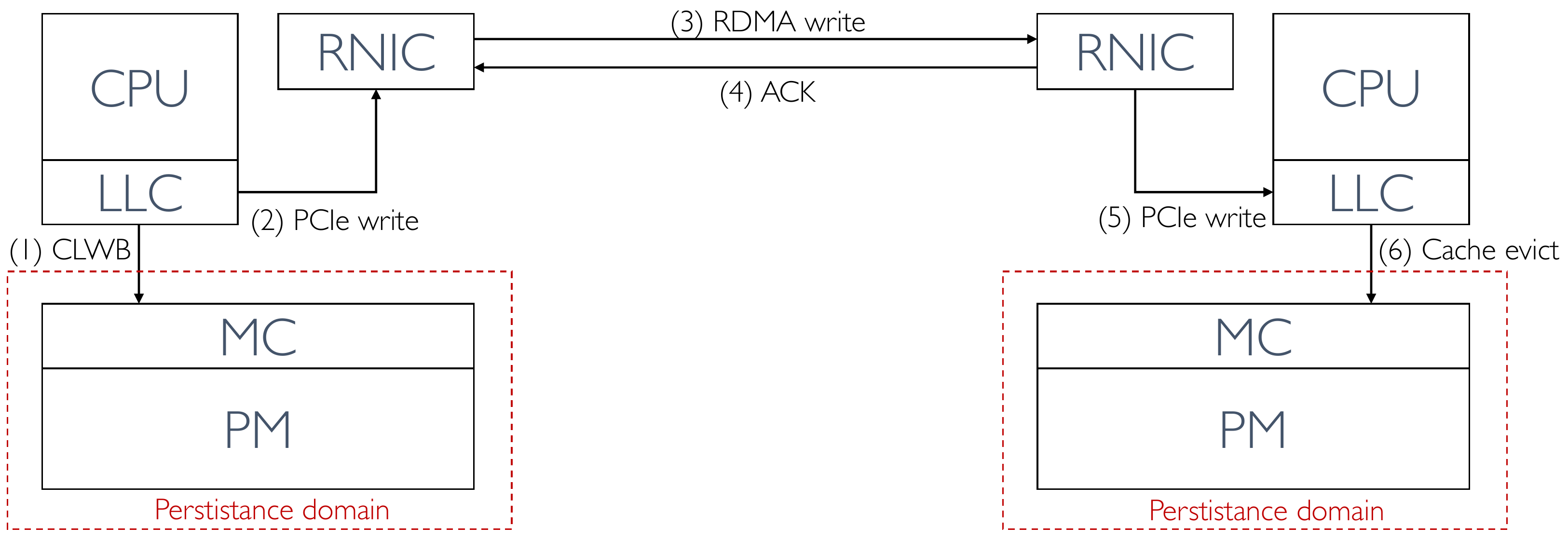}%
\caption{A primary-backup RDMA configuration. (1) Source flushes a cache-line to local PM; (2) source issues an RDMA write and busy waits (polls CQ); (3) RDMA write; (4) RDMA write ack (RDMA writes are posted); (5) PCIe write to backup LLC; (6) replicated cache line eventually gets evicted to persistent memory (PM).}
\label{fig:storage_vs_memory}
\end{figure*}

In this paper, we assume a system implementing Intel's recently proposed persistency model~\cite{intel_isa} to ensure \emph{local persistence} of data. 
The key instruction introduced as part of this persistency model is the \emph{cache line write back} ($clwb$) instruction, which writes back a cache line from the cache to the memory controller write queue.
We further assume that the system has support for Asynchronous DRAM Refresh (ADR), mandatory on Intel platforms~\cite{intel_adr,intel_wq}, which guarantees that all accepted writeback requests at the memory controller write queue will be drained to persistent memory in the event of a system failure~\cite{intel_wq,intel_adr}. 
Under this persistency model, two writes to addresses A and B are guaranteed to persist in order A < B, by using the following pseudo-assembly-code sequence:
\begin{equation}
str A;\;  clwb A;\; sfence;\; str B;
\label{eq:pmempseudocode}
\end{equation}
The combination of $clwb$ and $sfence$ above ensures that the store to A becomes persistent (by reaching the write queue at the memory controller) before the store to B.
And, this programming idiom can be used to provide more intuitive durability guarantees (e.g., transactions) to the regular programmers~\cite{kollipelley16}.
For processors which do not support the recently announced $clwb$ instruction, it can be replaced with the more expensive $clflush$ or $clflushopt$ instructions to provide the same persistence guarantees albeit at degraded performance~\cite{intel_isa}.

On top of achieving local persistence of data, to ensure memory-level SM, a programmer has to communicate two important pieces of information to the underlying replication system: (1) the dirty data that needs to be replicated and (2) specify the synchronization point in program where the replication of the identified dirty data has to be completed. For example, in storage-level SM, the file system captures all the necessary information and relays it to the replication system~\cite{Mojim-2015}. System calls like \emph{fwrite, fappend} indicate the dirty data that needs to be replicated while those like \emph{fsync, msync} specify the synchronization point in the program where the dirty data has to be replicated. Similarly, for memory-level SM, the memory persistency model employed helps capture this information.

For programs annotated with Intel's persistency model, the $clwb$ instruction (if unsupported, then $clflushopt$ or $clflush$) identifies the dirty cachelines that need to be persisted and the $sfence$ specifies the synchronization point by which the dirty cachelines have to be persisted. Based on such annotations, one could use 
RDMA writes to replicate dirty cachlines and enforce ordering and durability guarantees.
Using one-sided - RDMA writes is a preferred way of synchronous mirroring, as it only involves one server. Because of its one-sided nature, 
RDMA writes are cheaper in terms of latency and throughput compared to two-sided mechanisms, such as RPC.
In the following section, we expand on why it is challenging to use one-sided - RDMA
operations in SM.

\subsection{Challenges in mirroring transactions over RDMA}
While mirroring PM transactions, the replication system must be cognizant of the ordering and durability guarantees that must be provided within and at the end of the transaction respectively.

\textbf{Ordering challenges:} When mirroring using RDMA, there are many points in the path to the PM on the remote machine that can violate either the ordering or durability or both guarantees necessary to ensure the failure atomicity and durability of transactions.
For example, successive RDMA writes maybe placed in different Queue Pairs at the local NIC and may be processed in different order.
Or, even if the writes reach the remote NIC in the correct order and are processed in the correct order, modern architectures employ DDIO which causes the writes to be written to the LLC of the remote machine.
The LLC may then evict the corresponding cacheline in any order.
So, any mechanism that replicates transactions must ensure that the writes to remote PM persist in the correct order.

\textbf{Durability challenges:} Durability guarantees are also not easily provided when replicating PM transactions over RDMA.
When a durability fence is executed, the replication system must ensure that all the prior RDMA writes have been persisted in the remote PM.
However, the path for an RDMA write from the local machine to the remote PM consists of many buffers and caches that can cause the request to be buffered at any point (even in the local machine) and not be completed.
So, the replication system must deploy mechanisms which guarantee that upon the execution of \texttt{dfence}, the RDMA writes will in fact be persisted on a remote PM.

Both of these challenges stem from the fact that the acknowledgement to an RDMA write request (as shown in Figure~\ref{fig:storage_vs_memory}) is sent before the PCIe write is posted let alone the write becoming persistent.



\subsection{Key Goals of This Study}
This work aims to understand the performance implications of using RDMA in
SM of persistent memories.
We would like to understand whether RDMA is well-suited for SM of persistent memories and, if not, identify key performance issues and propose new techniques and hardware primitives that will make SM over RDMA more efficient. The two key questions this paper aims to answer:


\begin{itemize}
    \item What does it take to enforce ordering and durability guarantees in SM using
    the rcommit RDMA primitive? And, what is the performance impact of such a replication strategy?
    \item What are the other possible replication strategies? What primitives / hardware extensions do they require? And, how to reduce the performance overhead of SM using RDMA?
\end{itemize}



%% file: 04.strategies.tex
\label{sec:strategies}

\setbox0=\hbox{%
\begin{minipage}{0.9in}
\begin{lstlisting}[
basicstyle={\scriptsize\ttfamily},
identifierstyle={\color{black}},
tabsize=2,
language={},
numbersep=8pt,
numbers=left,
xleftmargin=0.5cm,frame=tlbr,framesep=2pt,framerule=0pt,
morekeywords ={class,run}
]
        (a) LOCAL
        ---------
        
Start Tx
    St A; CLWB A;
    
    St B; CLWB B;
    
    SFENCE;
    
    St C; CLWB C;
    
    St D; CLWB D;
    
    SFENCE;
End Tx

\end{lstlisting}
\end{minipage}
}
\savestack{\listingA}{\box0}

\setbox0=\hbox{%
\begin{minipage}{1.2in}
\begin{lstlisting}[
basicstyle={\scriptsize\ttfamily},
identifierstyle={\color{black}},
tabsize=2,
language={[AspectJ]Java},
numbersep=8pt,
numbers=left,
xleftmargin=0.5cm,frame=tlbr,framesep=2pt,framerule=0pt,
morekeywords ={class,run}
]
        (b) SM-RC
        ---------
        
Start Tx
    St A; CLWB A;
    RDMA Write A;
    St B; CLWB B;
    RDMA Write B;
    
    SFENCE; RCOMMIT;
    
    St C; CLWB C;
    RDMA Write C;
    St D; CLWB D;
    RDMA Write D;
    
    SFENCE; RCOMMIT
End Tx

\end{lstlisting}
\end{minipage}
}
\savestack{\listingB}{\box0}

\setbox0=\hbox{%
\begin{minipage}{2.1in}
\begin{lstlisting}[
basicstyle={\scriptsize\ttfamily},
identifierstyle={\color{black}},
tabsize=2,
language={[AspectJ]Java},
numbersep=8pt,
numbers=left,
xleftmargin=0.5cm,frame=tlbr,framesep=2pt,framerule=0pt,
morekeywords ={class,run}
]
        (c) SM-OB
        ---------
        
Start Tx
    St A; CLWB A;
    // Write Through Write
    RDMA Write (WT) A;
    St B; CLWB B;
    RDMA Write (WT) B;
    
    SFENCE; ROFENCE;
    
    St C; CLWB C;
    RDMA Write (WT) C;
    St D; CLWB D;
    RDMA Write (WT) D;
    
    SFENCE; RDFENCE;
End Tx

\end{lstlisting}
\end{minipage}
}
\savestack{\listingC}{\box0}

\setbox0=\hbox{%
\begin{minipage}{1.9in}
\begin{lstlisting}[
basicstyle={\scriptsize\ttfamily},
identifierstyle={\color{black}},
tabsize=2,
language={[AspectJ]Java},
numbersep=8pt,
numbers=left,
xleftmargin=0.5cm,frame=tlbr,framesep=2pt,framerule=0pt,
morekeywords ={class,run}
]
        (d) SM-DD
        ---------
        
Start Tx
    St A; CLWB A;
    // Non-Temporal Write
    RDMA Write (NT) A;
    St B; CLWB B;
    RDMA Write (NT) B;
    
    SFENCE;
    
    St C; CLWB C;
    RDMA Write (NT) C;
    St D; CLWB D;
    RDMA Write (NT) D;
    
    SFENCE; RDMA Read 0;
End Tx

\end{lstlisting}
\end{minipage}
}
\savestack{\listingD}{\box0}

\begin{table*}[h!]
\caption{The code transformations necessary to implement various replication strategies discussed in this section.}
\label{tab:code}
\begin{tabularx}{\textwidth}{|X|X|X|X|}
\hline
\stackinset{l}{}{t}{}{}{\listingA} &
\stackinset{l}{}{t}{}{}{\listingB} &
\stackinset{l}{}{t}{}{}{\listingC} &
\stackinset{l}{}{t}{}{}{\listingD} \\
\hline
\end{tabularx}

\end{table*}

\begin{figure*}[t] \centering
\includegraphics[width=0.85\linewidth]{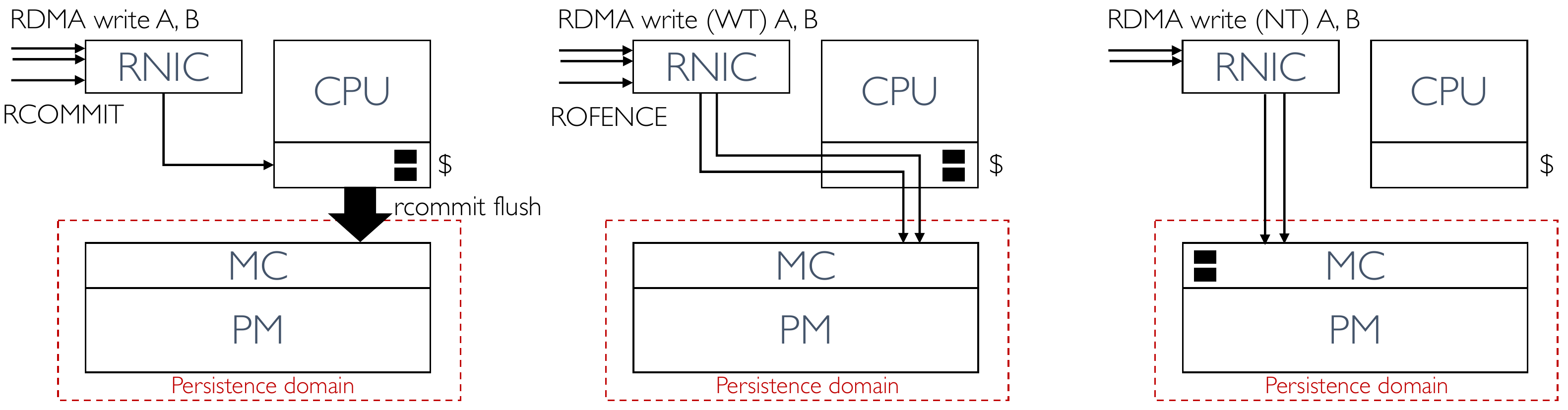}%
\caption{This figure depicts some of the new RDMA primitives proposed to enable and improve the performance of SM. (a) Depicts conventional RDMA writes and the new \rcommit primitive used in \smrc. (b) Depicts Write-Through writes (RDMA Write(WT)) that modify data in the LLC and then write back the data to the memory controller, a subsequent \rofence forces them to completion (c) Depicts Non-Temporal writes (RDMA Write(NT)) that directly write data to the memory controller bypassing the LLC.}
\label{fig:smtechniques}
\end{figure*}

%
%
%

In this section, we describe several strategies to synchronously mirror the contents of persistent memories and we evaluate each of these designs in subsequent sections.

\textbf{NO Synchronous Mirroring (\nosm):} \nosm is a replication strategy where we do not perform any replication but just persist data locally. 
While no credible replication strategy can realistically match the performance of \nosm, it provides a hypothetical upper bound on the possible performance and also helps us measure the costs introduced by various replication strategies.
Table~\ref{tab:code}(a) shows the code of a sample transaction with two epochs implemented using Intel's x86 extensions for PM programming (\clwb and \sfence).



\textbf{SM using Remote Commit (\smrc):}
Recently, there has been a push towards using single-sided RDMA verbs (like RDMA read() and write()) to avoid the overheads of interrupting the processor on the remote machine.
These read() and write() requests are completely handled by the receiving NIC without ever interrupting the associated CPU.
However, the challenge with using single-sided RDMA verbs for SM is that the completion of the write() verb does not provide any guarantees on whether the corresponding data persisted on the remote machine.
Upon receiving a completion acknowledgement, the sender of the write() message is just guaranteed that the write command has been communicated to the remote NIC.
The remote NIC may subsequently (at some future point) post a PCIe write command to write the corresponding data to the LLC of the attached processor (due to DDIO support).
Current PCIe interface does not provide any mechanisms to query if a posted write command has been completed.
Furthermore, even if the NIC can be assured that a posted PCIe write command is complete, the data has only been written to the LLC that is volatile and may lose its contents in the event of a system failure.
To achieve persistence, the corresponding data has to be written back from the LLC to the memory controller.


One way to achieve the persistence of data written to the LLC is to have a ``flusher'' thread running on the dedicated core in the processor and periodically interrupt the thread and request it to flush the data from the LLC.
However, that approach wastes valuable processor resources.
Taley and Pinkerton~\cite{rcommit} propose a single-sided RDMA verb called \rcommit to tackle this problem.
The completion acknowledgement of an \rcommit message guarantees to the sender that all prior issued RDMA writes are complete and that they are written back to the memory controller and hence persistent.
NICs that support the \rcommit primitive receive as input the range of memory addresses to be flushed to the persistent memory and have to ensure that all cachelines that belong to the specified range reach the memory controller before sending a completion acknowledgement. 
Figure~\ref{fig:smtechniques}(a) shows how an \rcommit writes back all the cachelines touched by prior writes from the cache hierarchy to the persistent memory controller.
We discuss the implementation details of \rcommit and how we model it in \S\ref{sec:methodology}.

\rcommit is both an ordering and durability primitive as the completion of an \rcommit ensures that all prior RDMA writes are durable.
That is, RDMA writes separated by an \rcommit will necessarily persist in order.
So, an \rcommit is an overloaded primitive (like an \sfence) that provides both ordering and durability guarantees.
Table~\ref{tab:code}(b) shows the code a sample transaction with two epochs being replicated using the SM-RC approach.
For each \clwb encountered, an RDMA write is issued and at the end of every epoch (indicated by an \sfence), an \rcommit is issued, whether its for intra-transaction ordering guarantees or for the end of transaction durability guarantees.



\textbf{SM using Ordered Buffering (\smob):}
In this design, we propose to decouple RDMA primitives for ordering and durability of remote writes.
We introduce a new primitive called \rofence, that ensures the ordering of remote writes, while a separate \rdfence primitive ensures their durability.
The \rofence demarcates different epochs within a transaction and hence allows the remote NIC to process the write requests in the correct order.
This approach allows all writes in the same epoch to be processed concurrently while writes in different epochs will be persisted in order.
While processing a write request, it is not enough for a remote NIC to simply post a PCIe command as that would result in the data being written to the LLC only.
We introduce a new PCIe command to allow the the data to not only be written to the LLC but also to be written back to memory controller.
We refer to this new PCIe command as a Write-Through Writes (Write (WT)).
Figure~\ref{fig:smtechniques}(b) highlights how the RDMA Write (WT) first is written to the LLC and then immediately also written through to the memory controller.
A subsequent \rofence ensures that all prior write through writes have been completed.
The durability fence, \rdfence, ensures that all outstanding RDMA writes and \rofence commands are executed before it completes.

Table~\ref{tab:code}(c) shows a sample transaction with two epochs being replicated using the \smob approach.
For every \clwb encountered, a RDMA write with Write-Through attribute is issued.
Further, an \rofence is issued at the end of each epoch within a transaction to achieve ordering guarantees while an \rdfence is issued at the end of the transaction to achieve durability guarantees.


\textbf{SM using Disabled DDIO (\smdd):}
\smob causes the RDMA writes to be first written to the LLC on the remote machine and then be written back to the memory controller.
In this approach, we disable the DDIO optimization on the remote machine allowing the remote writes to bypass the LLC and go directly to the memory controller.
Using this approach, the remote NIC can issue Non-Temporal Write requests (RDMA Write (NT)), that when completed are guaranteed to be persistent, as shown in Figure~\ref{fig:smtechniques}(c).
Furthermore, we route all the RDMA writes executed by the application through a single queue pair on the local and remote NICs.
By routing all the write requests through a single queue pair, we ensure that all the writes are processed by both the NICs (local and remote) in the order in which they were executed, ensuring the necessary ordering guarantees without the need for an explicit \rofence command.
And, to ensure the durability of all outstanding write commands, we issue a RDMA read command and poll for its completion. 
By the properties provided by the RDMA standard, the completion of the RDMA read ensures that all prior writes are completed and by disabling DDIO, we can be sure that the writes are indeed persistent.

Table~\ref{tab:code}(d) shows a sample transaction with two epochs being replicated using the \smdd approach.
For every \clwb encountered, a RDMA write with Non-Temporal attribute is issued, leading to bypass the LLC on the remote machine and directly reach the memory controller.
No fences are required to order writes from different epochs as \smdd provides an implicit guarantee that all writes will persist in program order (even those within the same epoch).
However, at the end of the transaction, to achieve durability guarantees, we need to ensure that all outstanding write requests are completed.
In order to achieve this durability guarantee, we issue an RDMA read to a sentinel address (0 in this case) which forces all prior writes to be completed (owing to the relationship between RDMA reads and writes).

\textbf{Discussion:} The main benefits of disabling DDIO for the purposes of SM are: (1) the RDMA writes are faster as they can now bypass the LLC and directly reach the persistent memory controller, (2) the overheads of \rofence, like additional RDMA commands incurred, are completely eliminated, and (3) designing the replication system is simpler as the designer is relieved of the burden of accurately identifying and placing the appropriate \rofence commands.
However, the downsides of this approach are: (1) all the RDMA writes issued for SM are routed through a single QP (for ordering purposes), failing to take advantage of the parallelism provided by multiple QPs at the NIC, and (2) today's computer systems have multiple tenants, so disabling DDIO for the benefit of one application might hurt the performance of other applications. 
While the first concern is fundamental to this approach, the second can be eliminated by a feature that can enable/disable DDIO on a per application or on an per QP basis.


While \smdd is implementable on certain select motherboards (which we unfortunately do not have access to), \smrc and \smob are not feasible on current systems as they require support for new RDMA verbs and PCIe commands.
To make a fair comparison of the different approaches, we develop an emulation methodology that we describe next and then present a comprehensive performance analysis of the different approaches.

%% file: 05.methodology.tex
\label{sec:methodology}

While SM for Persistent Memories is important, efficient mechanisms to enable SM are yet to be developed.
The mechanisms that we are proposing in this paper need new primitives that are not available in today's computing systems.
To estimate the performance impact of our proposed techniques, we model the latencies of the necessary remote memory operations that are currently not available.
To this end, we create a precise model of the last level cache (LLC) and the memory controller and investigate how local and remote operations interact with this model.
We use this model to measure delays when interacting with the memory subsystem. We combine these delays with actual RDMA network traffic to gain insights and compare end-to-end performance of SM for Persistent Memories using techniques proposed in this paper (\smob and \smdd) and the existing \smrc proposal.

We first describe our model of the last level cache and the memory controller before discussing how we estimated the performance of our new primitives.

\subsection{Modeling the LLC and the Memory Controller}

Given a certain physical address, we can find the LLC cache set in which it will be stored using existing models for Intel processors~\cite{10.1007/978-3-319-26362-5_3}. On the remote side, we measure that 10\% of each cache set in L3 is dedicated to DDIO traffic (2 out of the available 20 ways in the Xeon E5-2630 v3 processor in our test bed). Hence, assuming a least recently used (LRU) policy for DDIO traffic and the knowledge of physical addresses of the remote RDMA buffer, we can predict how each RDMA write will change the state of the remote cache and could potentially evict items from the LLC. 

Assuming a system with DDIO support, items that are evicted from the LLC will end up in the memory controller queue before being written to the persistent memory. Since the memory controller is part of the persistent domain, we need to ensure that the items enter this queue in the correct order. In systems without DDIO support, the writes from the network (i.e., from PCIe) will directly enter the memory controller's queue in the same order that they arrive. 

Note that once the memory controller's queue is full, the items cannot be inserted either from the LLC or the network. In this case, a back-pressure mechanism ensures that we can no longer send write requests to the remote server. Assuming a processor with a certain number of cache sets, our model takes four parameters as input: the latency of a PCIe write to the LLC (default - 200ns round-trip), the latency of a write from LLC to the memory controller's queue (10ns), the size of the memory controller's queue (64), and the latency of a write from the memory controller's queue to the persistent memory (150ns). Once configured, our model can estimate the latency of a given memory write at a certain address with a given size.

\subsection{Modeling Persistent Memory Primitives}

We now describe the individual remote and local memory access operations and how we model them.

\textbf{Cacheline Writeback (CLWB).} The \clwb instruction writes back dirty cachelines from the cache hierarchy to the memory controller. 
The instruction initiates a writeback and the programmers have to use a subsequent \sfence to ensure the completion of all prior \clwb instructions.

The \clwb instruction is going to be available in the Skylake server micro-architecture. To model \clwb, we rely on our model tag the cachelines that should be written to the memory controller's queue eagerly. At each \clwb, we calculate how many previously tagged cachelines have had time to have been eagerly sent to the memory controller. To model \sfence, we calculate how many tagged cachelines still need to be written to the memory controller and return the necessary delay for these writes to be completed. 

\textbf{Remote Commit (RCOMMIT).} \rcommit is a new RDMA verb being proposed in the RDMA standard to writeback the cachelines touched by all prior RDMA write operations from the LLC to the memory controller on the remote machine. The completion of an \rcommit indicates that all prior RDMA writes have been persisted on the remote machine.

To estimate the latency of an \rcommit, we treat is as being two separate operations performed in series.
First, the round trip to the remote machine, that involves setting up the QP entry, draining it on the local RNIC, receiving it on the remote RNIC, issuing the appropriate PCIe command and finally sending a response back. Second, the actions performed on the remote machine, that is the write back operations.
Hence, we model an \rcommit as being two operations performed in series, a remote RDMA read (to estimate the latency of a round trip) and the price of draining the previously written cachelines that are still present in the LLC as estimated by our model. This also means that some RDMA write operations may observe increased latency if the target cache set and the memory controller's queue are both full.

\textbf{Remote Writes Write-Through (\rwtw) and Non-Temporal (\rntw).} \rntw and \rwtw are new RDMA primitives we introduce to write the cachelines directly to the memory controller instead of just writing them to the LLC. When an application issues these new RDMA verbs and it is received on the remote RNIC, the remote RNIC tags the cachelines associated with this write to be either directly sent to the memory controller bypassing the LLC (\rntw) or written through from the LLC to the memory controller (\rwtw) from the PCIe root complex. 

To estimate the latency of an \rntw or \rwtw, we treat it as being two separate operations performed in series. First, the round trip to the remote machine, that involves setting up the queue pair entry, draining it on the local NIC, receiving it on the remote NIC, issuing the appropriate PCIe command and finally sending a response back. Second, the action performed on the remote machine. For the non-temporal write (\rntw), we use our model to estimate the latency assuming no DDIO support (i.e., writes directly go to the memory controller's queue). For the write-through write (\rwtw), we use the same estimation from our model as \clwb.



\textbf{Remote Ordering Fence (\rofence).} \rofence defines ordering points within the execution of an application where the hardware must ensure that all prior issued RDMA writes or \rntw or \rwtw are guaranteed to persist before any subsequent RDMA writes.
The exact mechanism by which the hardware provides this guarantee is left up to the discretion of the hardware designer.
One possible implementation is that the remote NIC receiving the RDMA write 
and \rofence commands, places them in a single FIFO queue.

To estimate the latency of an \rofence, we simply treat is an RDMA write which incurs a standard round trip latency.
The reason for this choice is that the \rofence, like an RDMA write, simply notifies the remote NIC of certain action to be performed (the write for RDMA write, while the ordering specification in case of \rofence) and does not actually force the completion of any action on the remote machine.

Such a queue introduces two sources of overheads: (1) it serializes the commands that are received from multiple independent threads/applications, (2) it does not allow any reordering of RDMA writes between successive \rofence commands.
Hardware designers may implement optimized versions of the queue that mitigates one or both of the overheads described.
Note that \rofence does not change the state of the LLC and/or the memory controller's queue and hence need not to interact with our model.

\textbf{Remote Durability Fence (\rdfence).} \rdfence defines the durability points within the execution of an application where the hardware must ensure that all prior issued RDMA writes 
are guaranteed to persist before the completion of the \rdfence command.
Note that while persisting data due to an \rdfence command, the hardware must still provide the ordering guarantees specified by any prior \rofence commands.

To estimate the latency of \rdfence command, we treat it as two separate serial operations, a round trip and draining writes to the persistent memory, similar to an \rcommit command. The notable exception for \rdfence is that the cachelines are eagerly written to the memory controller's queue from the LLC, following the same mechanism as \clwb. The latency of the writes to the persistent memory is hence estimated by our model similar to \clwb.  


\subsection{Experimental Setup}
\label{section:exp_setup}
To evaluate memory-level SM on RDMA, we build an RDMA test-bed based on Infiniband. Our test-bed assumes the following server configuration; an Intel Xeon E5-2630 v3 CPUs running at 2.4GHz and DDR4 memory. Our platform is based on Mellanox ConnectX-3 40Gbps RDMA network adapters. The IB platform is equipped with Infiniband SX6036 switch.
A detailed summary of our platform is presented in Table~\ref{table:platforms}.

\begin{table}[h!] 
\caption{RDMA platform}
\centering 
\renewcommand{\arraystretch}{1.4} 

\begin{tabular}{|m{1.5cm} m{5cm}|}
\hline 
\textbf{Component} & \textbf{Platform}\\
\hline 
CPU & Intel Xeon E5-2630 (2.4GHz)\\ 
\hline
Memory & DDR5 - 2400MHz (16x16GB DIMMs)\\
\hline
Network & Infiniband (40Gbps)\\
\hline 
RNIC & Mellanox ConnectX-3\\ 
\hline 
Switch & SX6036 (w/FDR)\\
\hline
\end{tabular}
\label{table:platforms}
\end{table}



\textbf{Platform disclaimers:} Our platform does not support the newly introduced $clwb$ or $clflushopt$ instructions, so we use the $clflush$ instruction in our programs. 
The RDMA write verb, once completed on current hardware does not guarantee that the data has been persisted to the remote memory, instead, it just guarantees that a PCIe write for the data has been posted in the remote machine.
This problem is already being addressed by the industry~\cite{SNIA-White-Paper-2016} and we expect future RNICs to close this vulnerability.

%% file: 06.real_applications.tex
\label{section:applications}
To compare the performance of the different replication strategies, we instrument a microbenchmark and workloads from the WHISPER~\cite{nalliharia17} with the primitives detailed in \S\ref{sec:strategies} and execute them on our test-bed with our model.

\subsection{Microbenchmarks}
\label{subsection:microbenchmarks}

\begin{figure*}[!t]
  \centering
  \includegraphics[angle=0,width=0.95\textwidth]{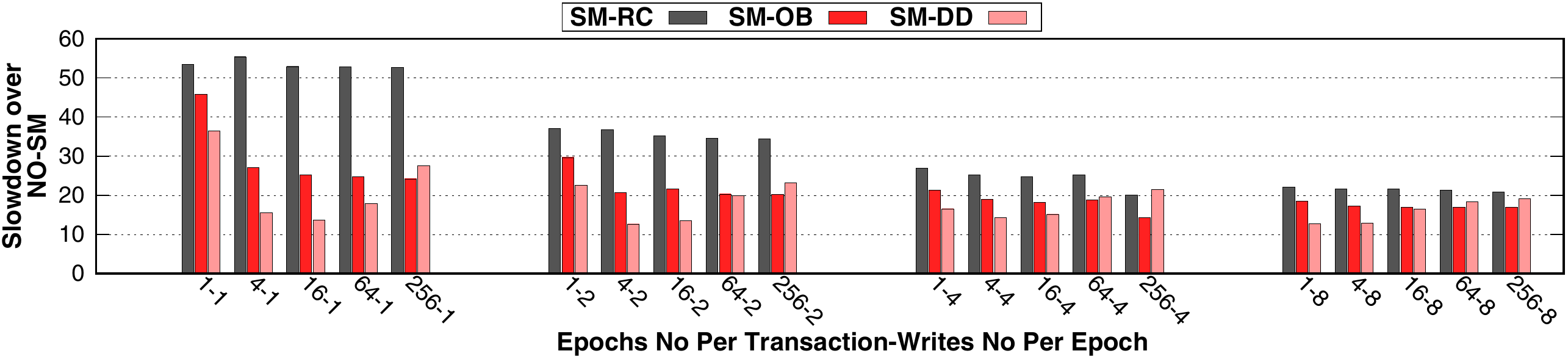}\vspace{-1mm}
  \caption{Slowdowns seen due to SM of Transact with different replication strategies over a baseline of \nosm. The x-axis labels are of the form $e-w$, where $e$ represents the number of epochs per transaction and $w$ represents the number of writes per epoch used in Transact.}
\label{fig:slowdowns}
\end{figure*}

Our microbenchmark, \emph{Transact}, executes 1M transactions, each of which is replicated.
Transact allows configurable number of epochs per transaction and further, configurable number of writes per epoch.
However, once configured, all the transactions have the same number of epochs per transactions and writes per epoch.
We vary the number of writes per epoch from [1 ... 8] and the number of epochs per transaction from [1 ... 256].
The ranges were decided based on the analysis of persistent memory workloads presented by Nalli \etal~\cite{nalliharia17}, which showed that applications exhibit very few writes per epoch, but can exhibit hundreds of epochs per transaction.
The addresses of writes within a transaction are randomly chosen. 

Our goal with this study is to understand the suitability of different replication strategies to different kinds of transactions - small transactions, large transactions, ones with fewer epochs, etc.
We evaluate Transact on the platform described in Section~\ref{section:exp_setup}.
Figure~\ref{fig:slowdowns} shows the slowdowns incurred with different replication strategies over a baseline of \nosm, where no replication is performed. 
The x-axis labels are of the form $e-w$, where $e$ represents the number of epochs per transaction and $w$ represents the number of writes per epoch used in Transact.
For example, a configuration of $16-2$ indicates that there are $16$ epochs per transaction and $2$ writes per epoch.
We make three major observations from Figure~\ref{fig:slowdowns}.







\textbf{First, \smrc incurs the highest performance overheads.} 
As Figure~\ref{fig:slowdowns} shows, \smrc incurs the highest slowdowns for all different Transact configurations, ranging from about 20x to over 55x.
\smrc uses expensive \rcommit verbs to ensure ordering and durability guarantees and the frequent use of this primitive results in the high overheads.
These overheads are highest for transactions with only 1 write per epoch and gradually reduce with increasing number of writes per epoch.
This behavior is to be expected as higher number of writes per epoch allows the an expensive \rcommit verb to be amortized over a larger number of asynchronous RDMA writes.

\textbf{Second, \smob and \smdd consistently outperform \smrc.}
Both \smob and \smdd outperform \smrc for all Transact configurations by as much as 3.5x (for Transact 4-1).
The main reason for the performance advantage of \smob and \smdd is that they eliminate the use of blocking \rcommit verbs to ensure intra-transaction ordering guarantees.
The performance difference is especially high when Transact has a low number of writes per epoch, due to a high ratio of \rcommit verbs per write.

\textbf{Third, \smob and \smdd are suitable to different kinds of transactions.}
Controlling for the number of writes per epoch within a transaction, \smob performs better for transactions with higher number of epochs per transaction while \smdd performs better for transactions with lower number of epochs per transactions.
This trade-off in performance with \smdd and \smob is to be expected as they optimize different aspects of SM.
For example, \smdd reduces the number of blocking ordering enforcement points than \smob by eliminating the need for \rofence primitive.
However, \smdd increases the latency of individual RDMA write operations by forcing each write to be written directly to the memory controller.
Furthermore, \smdd may cause frequent pauses to write processing at the remote NIC as it might cause the small write queue (64 entries, 4KB) at the memory controller to fill up while \smob allows the remote NIC to process more requests to buffer more writes in the LLC (up to 2MB as described in Section~\ref{sec:methodology}).
For these reasons \smdd is more efficient for small transactions while \smob is more efficient for larger transactions.

\textbf{Discussion.} With the optimized cache flush instructions ($clflushopt$ and $clwb$), we expect the time to perform local flush operations to reduce significantly providing fewer opportunities to overlap remote RDMA writes and local flush operations resulting in higher performance overheads even for larger transactions.
However, Transact is dominated by writes and real applications 
with less frequent write operations can expect lower overheads. 
Next, we analyze the overheads experienced with real applications from the WHISPER~\cite{nalliharia17} benchmark suite.

\subsection{WHISPER applications}
\label{subsection:benchmarks}

\begin{figure}[!t]
  \centering
\subfloat[Execution Time]{\label{fig:micro_slowdown_ib}\includegraphics[angle=90,width=0.8\columnwidth]{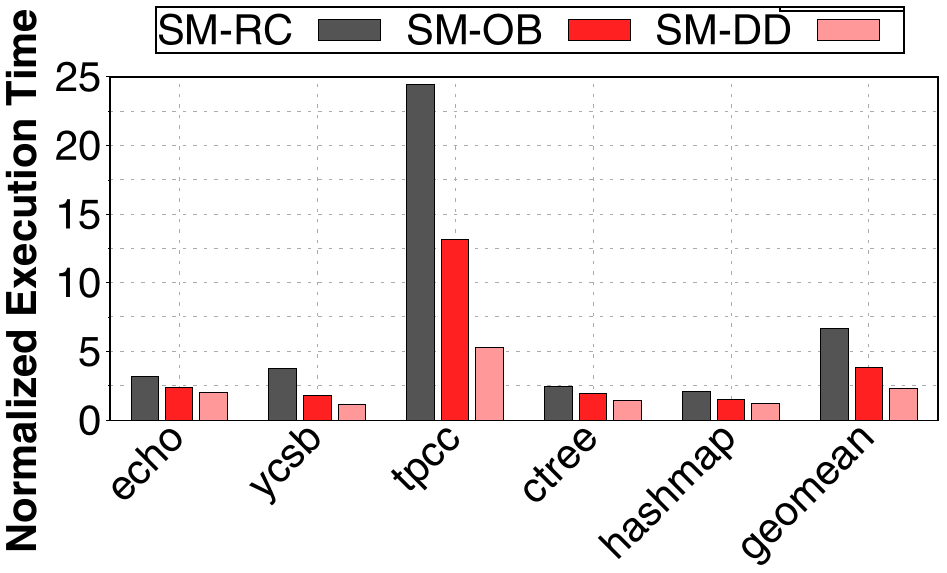}}\vspace{-4mm}\\

\subfloat[Throughput]{\label{fig:micro_slowdown_ib_tp}\includegraphics[angle=90,width=0.8\columnwidth]{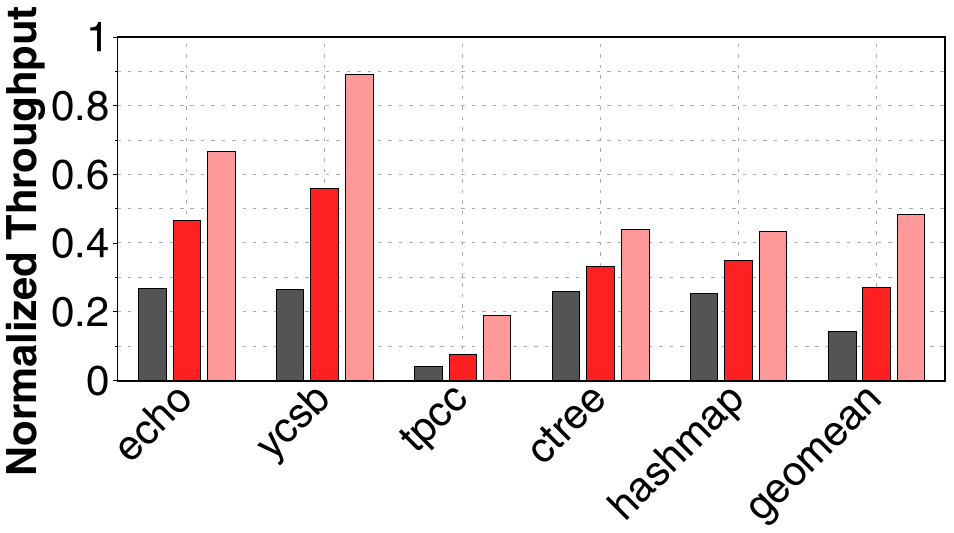}}\vspace{-2mm}\\
  \caption{Changes in the execution time and throughput due to SM for different WHISPER workloads.}
\label{fig:WHISPER-slowdowns}
\end{figure}

In this subsection, we first present a quick overview of persistent memory applications from the WHISPER~\cite{nalliharia17} benchmark suite and then show the overheads of employing memory-level SM for these workloads.

\noindent \textbf{C-tree} is a multi-threaded benchmark that performs inserts and deletes on a persistent crit-bit tree data structure. This workload was initially released as part of NVML~\cite{nvml}.

\noindent \textbf{Echo} mimics the behavior of a persistent key-value store. The benchmark consists of a master thread that manages the store and many client threads that send updates to the store.

\noindent \textbf{Hashmap} is a multi-threaded benchmark that performs inserts and deletes on a persistent crit-bit tree data structure. This workload was initially released as part of NVML~\cite{nvml}.

\noindent \textbf{YCSB and TPCC} are two transaction processing workloads operating over N-store, a relational database management system designed from scratch for persistent memories.

Next, we delve in the performance of different replication strategies for these workloads.



\textbf{\smdd $>$ \smob $>$ \smrc.}
Figure~\ref{fig:WHISPER-slowdowns} highlights the increased execution times and decreased throughputs for the WHISPER applications described above for various replication strategies.
As with Section~\ref{subsection:microbenchmarks}, the measurements presented for each platform are normalized over \nosm. We draw two key observations.
First, \smrc, on average, increases execution time by 6.7x, while \smob and \smdd increase execution times by only 3.8x and 2.3x, respectively. Hence, \smob and \smdd outperform \smrc by 1.8x and 2.9x, respectively. Second,
\smrc decreases throughput by 85\%, while \smob and \smdd decrease throughput by only 73\% and 52\%, respectively.
\smdd consistently performs best, while \smrc performs worst.

\textbf{Workload characteristics.} The relative performance of \smrc, \smob, and \smdd are to be expected based on the characteristics of the workloads.
These benchmarks have only a modest fraction of all the write instructions executed during the applications and update persistent data (about 5\%)~\cite{nalliharia17}.
We expect that the performance degradation due to memory-level SM for different applications will be largely determined by the fraction of accesses to persistent data, with applications having higher fractions of accesses to persistent data seeing more performance degradation.
The rest of the write instructions in the application arise due to modifications to volatile data structures not necessary for crash consistency.
Furthermore, the average number of writes per epoch for these workloads is only 1.4.
A disproportionate number of epochs in most of the workloads modify only one cacheline worth of data. 
However, the average number of epochs per transaction vary significantly for the various workloads from about 10 (hashmap) to over 300 (echo).
Based on the low number of writes per epoch in these workloads and our analysis of the microbenchmark Transact, we can expect \smdd to perform best.




%% file: 07.related_work.tex
\label{section:RelatedWork}

In this section, we present a brief overview of the research most relevant to our work.

\noindent\textbf{RDMA Applications.} The use of RDMA has gained popularity as a communication mechanism in many distributed database systems and algorithms. In the context of join algorithms, the costs of large data transfers can be significantly reduced by interleaving compute and communication, and by avoiding intermediate copies through the use of one-sided memory operations~\cite{barthels-sigmod-2015}. These algorithms have been shown to scale to several thousands of cores and are able to achieve a high throughput~\cite{barthels-pvldb-2017}. 
RDMA is used in BatchDB as a low-latency communication mechanism to replicate data to multiple workload-specific replicas in order to maintain strict performance, data freshness, and consistency requirements~\cite{makreshanski-sigmod-2017}.
Many distributed database systems have explored RDMA communication primitives as part of their query pipeline in order to speed up query response times~\cite{roediger-vldb-2017, kraska-ieee-2017}, while others have been re-designed from the ground up with RDMA in mind~\cite{binnig-vldb-2017, barthels-ieee-2017}.
Key-value storage systems make use of high-performance networks to enable low-latency access to remote data~\cite{farm-nsdi-2014,HERD-SIGCOMM-2014}.




\noindent \textbf{Persistent memory systems.} Emerging persistent memory technologies present a paradigm shift in storage technologies and have set off an assortment of research projects in the areas of persistent data structures~\cite{izraelevitzmendes16,nawabizraelevitz17,venkataramantolia11}, file systems~\cite{PMM-SOSP-2009,dulloor2015systems,PMM-SC-2011,volosnalli14,xuswanson16,aguilera2018remote}, transaction logging techniques~\cite{arulrajperron16,pelleywenisch13,chatzistergioucintra15,wangjohnson14,PMM-ICCD-2014,kollipelley16} and many more.
Many recent works, both from academia and industry, have also sought to develop new programming models~\cite{intel_isa,pelleychen14,joshinagarajan15,nalliharia17} to program for persistent memory systems and have also developed architectural solutions to efficiently implement these models~\cite{kollirosen16,joshinagarajan15,nalliharia17}.
However, all of these works assume that data is ``persistent'' as soon as it reaches local persistent memory, as assumption that cannot be employed in large scale multi-node systems.

\noindent \textbf{Replicating persistent memory data.}
Replicating data to ensure high-availability and high-durability is a widely used approach~\cite{adyabolosky02,calderwang11,decandiahatorun07,ghemawatgobioff03,grayhelland96,gibsonhellerstein88}.
While many different replication techniques are used, synchronous mirroring (SM) is a common feature of enterprise storage solutions~\cite{emc,hpe,netapp,vsan,ibm}.
For conventional storage and network technologies, the strong consistency guarantees of SM comes at a very low performance cost~\cite{maccormickmurphy04}.
However, with the emergence of persistent memory technologies a reevaluation of the costs of SM have to be reevaluated.

Recent proposals from the Storage Networking Industry Association (SNIA)~\cite{SNIA-White-Paper-2016,snia1} indicate that both hardware manufacturers and software developers are looking to enable efficient remote persistent memory access over RDMA. 
Mojim~\cite{Mojim-2015} enables SM for persistent memories and most closely resembles the systems envisioned in our work.
However, one major difference is that Mojim is designed to provide SM for file systems mounted on persistent memories and not designed to support the newer persistent memory programming interfaces~\cite{nvmpm}. 
Unlike the file systems, these newer interfaces expose a byte-addressable storage media to the programmers, resulting in fine-grain accesses (a few cachelines at a time~\cite{nalliharia17}).

%% file: 08.conclusion.tex
\label{section:conclusion}
In this work we propose a set of new RDMA primitives that aim to 
improve the efficiency of persistent memory mirroring. On top of that, we propose two novel SM techniques that leverage the new primitives to unleash the full
potential of modern RDMA hardware. The first technique is based on ordered buffering, while the other one uses non-temporal remote writes to ensure 
ordering and durability guarantees. We add the support for SM in the WHISPER benchmark suite and use a microbenchmark to evaluate the performance of 
our two proposals and compare them to a prior proposal based on the remote commit RDMA verb. We conclude that ordered buffering and non-temporal writes
enable highly concurrent SM and outperform designs that use remote commit.